\documentclass[aps,prb,twocolumn, showpacs,superscriptaddress,amsmath,amssymb]{revtex4}
\usepackage{graphicx}
\usepackage{amsmath}
\usepackage{textcomp}
\usepackage{graphicx}
\usepackage{dcolumn}
\usepackage{bm} 
\usepackage{color}
\usepackage{ulem}

\begin{document}

\title{%
Magnetic form factor analysis on detwinned single crystal of BaFe$_2$As$_2$}
\author{K.~Kodama}
\affiliation{Quantum Beam Science Center, Japan Atomic Energy Agency, Tokai, Ibaraki 319-1195, Japan }
\author{M.~Ishikado}
\affiliation{Comprehensive Research Organization for Science and Society (CROSS), Tokai, Ibaraki 319-1106, Japan }
\author{S.~Wakimoto}
\affiliation{Quantum Beam Science Center, Japan Atomic Energy Agency, Tokai, Ibaraki 319-1195, Japan }
\author{K.~Kihou}
\affiliation{Nanoelectronics Research Institute, National Institute of Advanced
   Industrial Science and Technology, Tsukuba, Ibaraki 305-8562, Japan }
\author{C. H.~Lee}
\affiliation{Nanoelectronics Research Institute, National Institute of Advanced
   Industrial Science and Technology, Tsukuba, Ibaraki 305-8562, Japan }
\author{A.~Iyo}
\affiliation{Nanoelectronics Research Institute, National Institute of Advanced
   Industrial Science and Technology, Tsukuba, Ibaraki 305-8562, Japan }
\author{H.~Eisaki}
\affiliation{Nanoelectronics Research Institute, National Institute of Advanced
   Industrial Science and Technology, Tsukuba, Ibaraki 305-8562, Japan }
\author{S.~Shamoto}
\affiliation{Quantum Beam Science Center, Japan Atomic Energy Agency, Tokai, Ibaraki 319-1195, Japan }


\date{\today}

\begin{abstract}
We have performed neutron diffraction measurement on a single crystal of parent compound of iron-based superconductor, BaFe$_2$As$_2$ 
at 12~K. In order to investigate in-plane anisotropy of magnetic form factor in the antiferromagnetic phase, 
the detwinned single crystal is used in the measurement.  
The magnetic structure factor and magnetic form factor are well explained by the spin densities consisting of $3d_{yz}$ electrons with a fraction of about 40~\% and the electrons in the other four $3d$ orbitals with each fraction of about 15~\%.  Such anisotropic magnetic form factor is qualitatively consistent with the anisotropic magnetic behaviors observed in the antiferromagnetic phase of the parent compound of 
iron-based superconductor.
\end{abstract}

\maketitle

In-plane anisotropy observed in electronic properties of iron-based superconductors is one of the important issues 
in relation to the mechanism of the superconductivity.  
The parent compound of the iron-based superconductor exhibits structural phase transition from tetragonal to 
orthorhombic structures with decreasing temperature, whereas the superconducting phase achieved by partial atomic 
substitution remains tetragonal structure down to the lowest temperature.  
In the superconducting phase and the tetragonal phase of the parent compound, the electronic properties exhibit two-fold 
symmetry in FeAs layer although the crystal structure is the tetragonal with four-fold symmetry.\cite{kasahara}  
Such in-plane anisotropy which is so-called nematic state probably originate from degrees of freedom of electrons.  
So-called spin nematic state and the orbital ordering are suggested as the origin of the strong breaking 
of four-fold symmetry.\cite{fernandes1, kruger, lv, lee, fernandes2}   
\par   
In the orthorhombic phases of a parent compound and non-superconducting underdoped compounds, 
the clear in-plane anisotropy has been observed in both of the electronic and magnetic properties.  
Electric resistivity along $b$-direction is about twice larger than the resistivity along $a$-direction 
in the orthorhombic phase of underdoped Ba(Fe$_{1-x}$Co$_x$)$_2$As$_2$.\cite{chu1, ishida} 
Optical conductivity below about 0.2 cm$^{-1}$ along $a$-direction is also about twice larger than  
the conductivity along $b$-direction in BaFe$_2$As$_2$.\cite{nakajima1, nakajima2}  
Drastic change of Fermi surfaces with from fourfold symmetry to that with twofold symmetry is observed below the antiferromagnetic transition temperature due to resolving a degeneracy between  $3d_{yz}$ and  $3d_{zx}$ orbitals.\cite{shimojima}
The energy dispersion of the spin wave in the antiferromagnetic phase observed by 
inelastic neutron scattering has large anisotropy which is explained by considering the nearest neighboring magnetic 
interactions along $a$ and $b$ directions with opposite sign.\cite{zhao, ewings, harriger}  
These anisotropic behaviors observed in the orthorhombic phase are more pronounced than that simply expected from the 
small difference between $a$ and $b$ lattice constants (less than 1~\%).  
These results suggest that the electronic state, for example, the spatial distribution of the $3d$ electrons 
which contribute to the electronic and magnetic properties is largely anisotropic.  
However, early neutron diffraction measurement on a single crystal sample of the parent compound including twinned domains 
has reported that the magnetic form factor which is estimated from magnetic Bragg intensities of $h0l$ reflections is 
almost isotropic,\cite{ratcliff} inconsistent with the above anisotropic magnetic behaviors.
In this letter, results of neutron diffraction measurement on a detwinned single crystal of BaFe$_2$As$_2$ in the 
antiferromagnetic phase are reported.  The magnetic form factor determined from the magnetic Bragg intensities of both of $hkh$ and $h0l$ reflections 
is anisotropic in $a$-$b$ plane, qualitatively consistent with the anisotropic magnetic behaviors. 
\par
A single crystal of BaFe$_2$As$_2$ was grown by self-flux method. The details are reported in ref. 16.  
The crystal with a volume of 4 $\times$ 4$\times$ 0.5 mm$^3$ was used in neutron diffraction measurement.  
The neutron diffraction measurement was performed using the triple-axis spectrometer TAS-1 installed at the research reactor 
JRR-3 of Japan Atomic Energy Agency.  The single crystal was detwinned in a sample holder made of Al 
by uniaxal pressure along $b$-axis in the orthorhombic phase.  
In the estimations of nuclear and magnetic Bragg intensities, the neutron absorption of the sample folder is corrected by considering the neutron flight paths in the sample folder for each Bragg reflections. 
The sample was sealed in an aluminum can, and then mounted in a closed-cycle He gas refrigerator.  
Collimation sequence of open-open-S-80$'$-open (S denotes sample) were used.  
Pyrolytic graphites (PGs) were used as a monochromator and an analyzer. 
Another PG was placed downstream of the sample to eliminate higher order neutrons.  
The detwinning was confirmed at 12~K by $\theta-2\theta$ scan  with the neutron wave length of 2.3532~$\mathrm{\AA}$, as shown in Fig. 1.
The nuclear and magnetic Bragg reflections were collected by $\theta-2\theta$ scan with the neutron wave length of 1.6377~$\mathrm{\AA}$. 
The crystal was oriented with two types of configurations including the reciprocal lattice points of $hkh$ and $h0l$ in the horizontal scattering planes.   
\par
Figures 1(a) and 1(b) show the profiles of 040 and 400 nuclear Bragg reflections obtained in the conditions to collect  
the intensity of $hkh$ and $h0l$ reflections, respectively.  The data are collected at room temperature and 12~K where the 
crystal structures are tetragonal and orthorhombic, respectively.  Here, because we use the notation of the orthorhombic structure, the 
reflections at room temperature correspond with 220 reflection in the tetragonal structure.  
\begin{figure}[tbh]
\centering
\includegraphics[width=8cm]{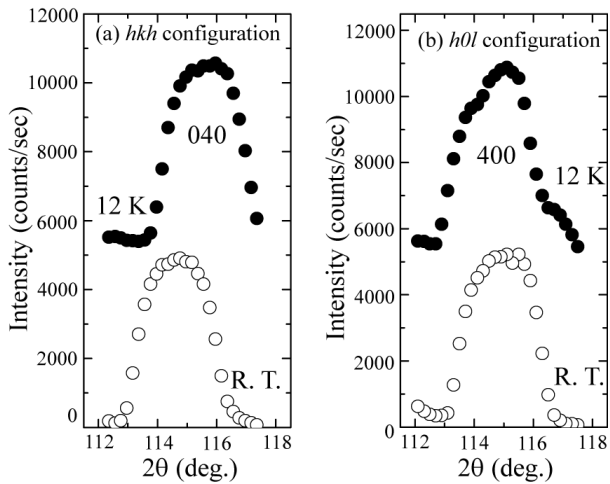} 
\caption{Peak profiles of 040 (a) and 400 (b) nuclear Bragg reflections obtained in the scattering conditions to collect  
the intensity of $hkh$ and $h0l$ reflections, respectively.  The data are collected at 12~K (closed circles) and room temperature (open circles). }
\label{fig.1}
\end{figure}
The peak width at room temperature almost corresponds with the instrumental resolution (momentum transfer of about 0.1~\AA$^{-1}$ in the condition with the neutron wave length of 2.3532~$\mathrm{\AA}$).  
040 and 400 reflections of the orthorhombic phase at 12~K are observed at different positions and their 
peak widths almost correspond with the width at room temperature, indicating that the detwinned single crystal is obtained at 12~K.  
The lattice constants at 12~K estimated from the positions of the peak centers of 400, 040 and 002 reflections are 
$a=5.601~\mathrm{\AA}$, $b=5.568~\mathrm{\AA}$ and $c=12.95~\mathrm{\AA}$, respectively. 
In Fig. 2, the observed nuclear structure factor squared, $\left| F_N\right|_{obs}^2$, is plotted against the calculated 
structure factor squared, $\left| F_N\right|_{cal}^2$.  
The observed data are corrected by Lorentz factor, $L(\theta)$.  
Here, we use the atomic positions reported for the orthorhombic phase in the calculation of 
$\left| F_N\right|_{cal}^2$.\cite{rotter} 
The solid line is the fitting result by using a formula 
$\left| F_N\right|_{obs}^2=A \left| F_N\right|_{cal}^2 (1-B \left| F_N\right|_{cal}^2)$, 
where $A$ is a scale factor and $B$ accounts for extinction.\cite{shamoto} 
\begin{figure}[tbh]
\centering
\includegraphics[width=7.0 cm]{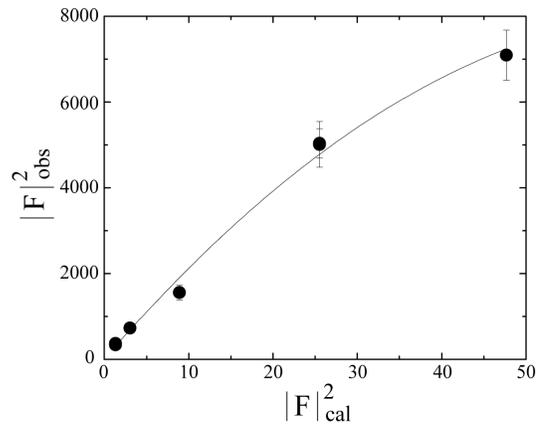} 
\caption{Observed nuclear structure factor squared, $\left| F_N\right|_{obs}^2$, is plotted against 
squared nuclear structure factor $\left| F_N\right|_{cal}^2$ calculated by using the atomic positions reported in ref. 17.
Solid line is fitting result by using a formula 
$\left| F_N\right|_{obs}^2=A \left| F_N\right|_{cal}^2 (1-B \left| F_N\right|_{cal}^2)$.}  
\label{fig.2}
\end{figure}
\par
By using the scale factor $A$ determined from the analysis on nuclear Bragg intensities, the intensities of 
magnetic Bragg reflections are described by using following equations. 
\begin{eqnarray}
I_{obs}(\bm{Q})&=&A\times L(\theta) \left| F\right|_{obs}^2,\\ \nonumber
\left| F\right|_{obs}^2&=&\gamma_0^2 \mu^2 f_{obs}(\bm{Q})^2 \\ 
 &\times& \left| \sum_n \sin\alpha_n \exp[(2\pi i(hx_n+ky_n+lz_n)] \right|^2, \\ 
L(\theta)&=&1/\sin 2\theta,
\end{eqnarray} 
where the summation is taken over all magnetic moment in the magnetic unit cell and $\gamma_0=0.269\times10^{-12}$ cm, 
$\mu$ is the amplitude of ordered magnetic moment in the unit of Bohr magneton, and $\alpha$ is the angle between the 
direction of $n$-th magnetic moment and the scattering vector $\bm{Q}$.  
The magnetic structure in the antiferromagnetic phase has already been reported, as schematically shown in Fig. 3(a).
\cite{zhao2, huang, goldman}
The magnetic moments have antiferromagnetic and ferromagentic arrangements along $a$- and $b$-directions, respectively, 
and are almost parallel to $a$-direction, which is so-called stripe type magnetic structure.  
For this magnetic structure, the magnetic reflection is observed at $hkl$ with $h=2n+1$, $k=2n$ and $l=2n+1$, 
where $n$ is an integer.  
If we assume that the direction of the magnetic moment is slightly away from $a$-axis in $a$-$b$ plane, $\phi$, as shown in Fig. 3(a), 
the magnetic structure factor depend on $\phi$ and eq. (2) can be rewritten as follows.  
\begin{eqnarray}
\nonumber \left| F\right|_{obs}^2&=&64 \gamma_0^2 \mu^2 f_{obs}(\bm{Q})^2 \\ 
&\times& \left| \sin\left\{\cos^{-1}\left(\frac{ha^*\cos\phi+kb^*\sin\phi}{Q \mu}\right)\right\} \right|^2.
\end{eqnarray} 
Although the observed magnetic structure factor squared, $\left| F\right|_{obs}^2$, and the observed magnetic form factor,
$f_{obs}(\bm{Q})$, can be determined from the observed magnetic Bragg intensities and the equations (1), (3) and (4), 
the estimated value of the latter depends on $\phi$.  
\begin{figure}[tbh]
\centering
\includegraphics[width=7.0 cm]{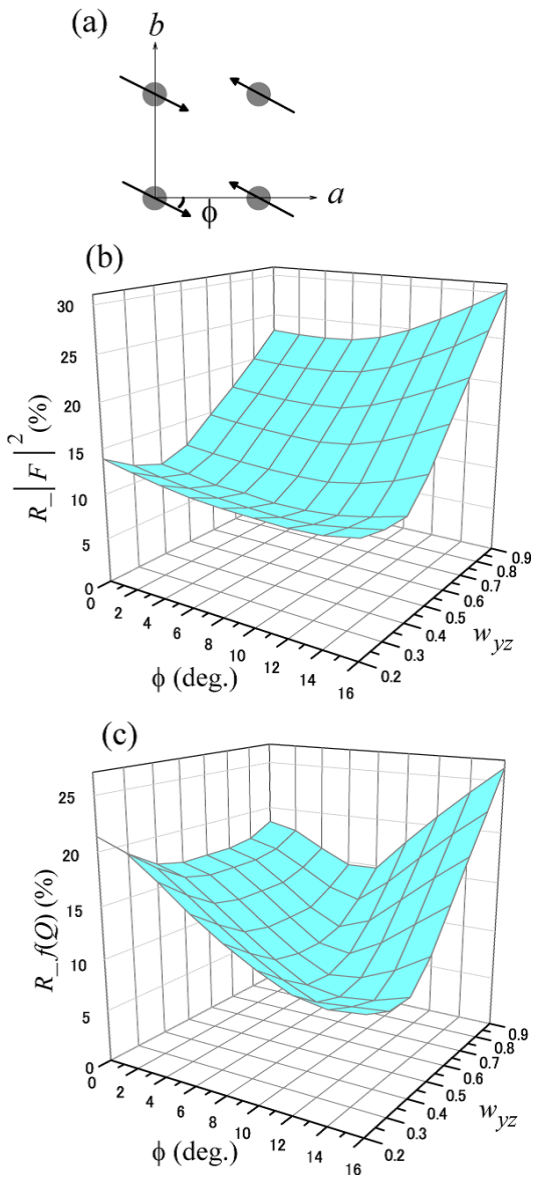} 
\caption{(Color online) (a) Magnetic structure in antiferromagnetic phase of BaFe$_2$As$_2$.  
$R$-factors estimated from $\left| F\right|_{obs}^2$ and $\left| F\right|_{cal}^2$ (b), and those from 
$f_{obs}(\bm{Q})$ and $f_{cal}(\bm{Q})$ (c), are plotted against $\phi$ and $w_{yz}$.  }
\label{fig.3}
\end{figure}
Here, we compare the $\left| F\right|_{obs}^2$ and $f_{obs}(\bm{Q})$ with the calculated magnetic structure factor squared, 
$\left| F\right|_{cal}^2$, and magnetic form factor, $f_{cal}(\bm{Q})$, which are obtained by following equations. 
\begin{eqnarray}
\nonumber \left| F\right|_{cal}^2=64 \gamma_0^2 \mu^2 f_{cal}(\bm{Q})^2 \\ 
\times \left| \sin\left\{\cos^{-1}\left(\frac{ha^*\cos\phi+kb^*\sin\phi}{Q \mu}\right)\right\} \right|^2, \\
\nonumber f_{cal}(\bm{Q})=w_{yz} f_{yz}(\bm{Q})+\frac{1-w_{yz}}{4} \\ 
\times \left\{f_{xy}(\bm{Q})+f_{zx}(\bm{Q})
+f_{x^2-y^2}(\bm{Q})+f_{3z^2-r^2}(\bm{Q}) \right\}
\end{eqnarray} 
Here, $x$, $y$, and $z$ directions correspond with $a$, $b$, and $c$ directions of the orthorhombic lattice, respectively.  
Although the isotropic magnetic form factor has been reported,\cite{ratcliff} we consider the anisotropic one 
which has larger weight of the 3$d_{yz}$ orbital than the other orbitals because the ferro-orbital ordering model 
in which the electrons in 3$d_{yz}$ orbital has large contribution to the magnetic moment is suggested.\cite{lee}  
In the calculations of above equations, 
the form factors and the wave functions of 3$d$ orbitals reported on Fe ion are used.\cite{freeman1, freeman2, watson} 
\par
Figures 3(b) and 3(c) show $R$-factors estimated from $\left| F\right|_{obs}^2$ and $\left| F\right|_{cal}^2$, 
and from $f_{obs}(\bm{Q})$ and $f_{cal}(\bm{Q})$, for various $\phi$ and $w_{yz}$, 
respectively. They are estimated by following equations.  
\begin{equation}
R=\frac{\sum_i \left| W_i^{obs}-W_i^{cal} \right| }{\sum_i W_i^{obs}} , 
\end{equation} 
where $W_i^{obs}$ and $W_i^{cal}$ are observed and calculated values of the squared magnetic structure factors 
and magnetic form factors.  $\sum_i$ means a summation taken over all observed magnetic Bragg reflections.  
The $R$-factor has minimum value of 9.4~\% at $\phi$=10~deg. and $w_{yz}=0.4$ for the squared magnetic structure factor, 
and the $R$-factor for the magnetic form factor has minimum value of 8.0~\% at $\phi$=12~deg. and $w_{yz}=0.4$.  
The values of  $\phi$ and $w_{yz}$ where the $R$ obtained from $\left| F\right|_{obs}^2$ and $\left| F\right|_{cal}^2$ is minimum, is almost consistent with the values where the $R$ obtained from $f_{obs}(\bm{Q})$ and $f_{cal}(\bm{Q})$ is minimum.  
The weight of 3$d_{yz}$ orbital is larger than the each weight of the other 3$d$ orbitals 
($w_{yz}$ is about 0.4 and the other weights are about 0.15).  
They indicate that the such anisotropic magnetic form factor must be considered to reproduce the observed magnetic structure 
factor and magnetic form factor.  
In Figs. 4(a) and 4(b), $\left| F\right|_{obs}^2$ and $f_{obs}(\bm{Q})$ are plotted against $\left| F\right|_{cal}^2$ 
and $f_{cal}(\bm{Q})$, respectively.
\begin{figure}[bth]
\centering
\includegraphics[width=6.0 cm]{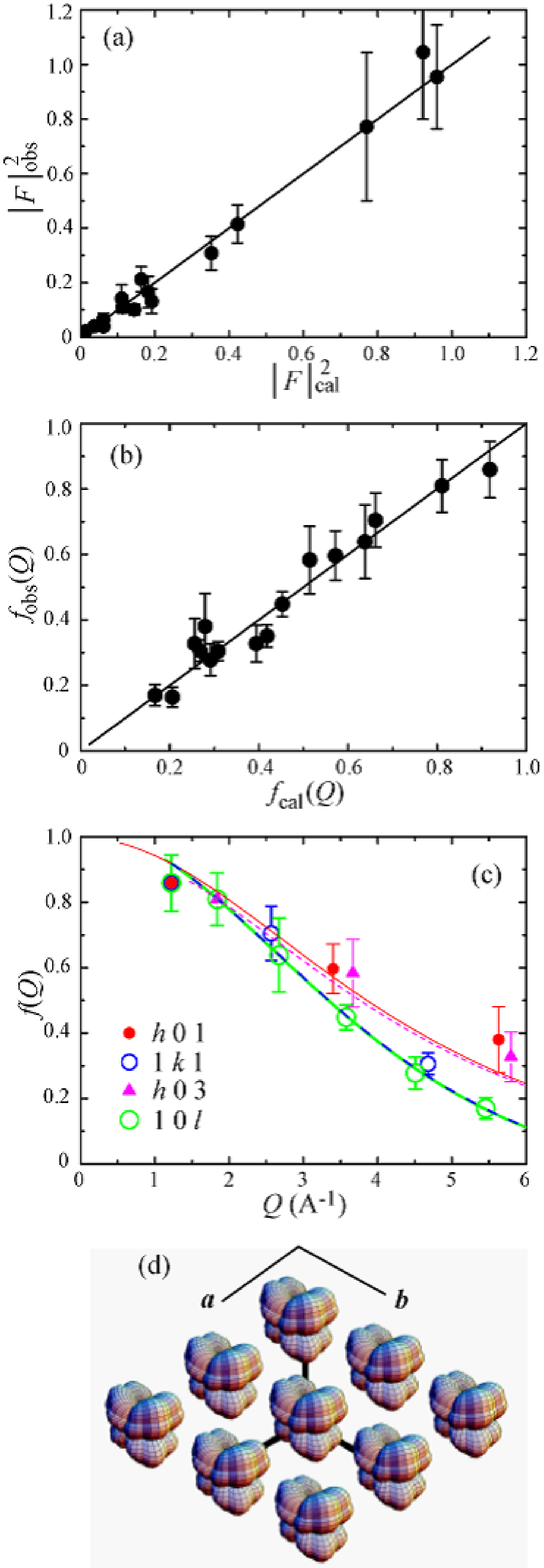} 
\caption{(Color online) $\left| F\right|_{obs}^2$ (a) and $f_{obs}(\bm{Q})$ (b) are plotted against 
$\left| F\right|_{cal}^2$ and $f_{cal}(\bm{Q})$, respectively.  Straight lines show $\left| F\right|_{obs}^2=
\left| F\right|_{cal}^2$ and $f_{obs}(\bm{Q})=f_{cal}(\bm{Q})$.  
(c) The $f_{obs}(\bm{Q})$ obtained at $h01$ (closed circles), $1k1$ (small open circles), $h03$ (closed triangles) 
and $10l$ (large open circles) reflections are plotted against $Q$.  
Thin solid and dashed lines are the magnetic forms factors calculated at $h01$ and $h03$, bold dashed and solid lines 
are the magnetic form factors calculated at $1k1$ and $10l$, respectively.  The bold dashed line corresponds with the 
bold solid lines.  (d) Spin density of Fe ions in the orthorhombic unit cell is schematically shown in $a$-$b$ plane.}
\label{fig.4}
\end{figure}
The plotted values are obtained for $\phi$=10~deg. and $w_{yz}=0.4$.  Here, we adopt the amplitude of magnetic moment, 
$\mu=0.70\mu_B$, optimized in the above analysis.  
The observed values are almost reproduced by the calculated values, and the weighted $R$ factors are 9.4~\% and 8.4~\% 
for the squared magnetic structure factor and the magnetic form factor, respectively.  
In Fig. 4(c), $Q$-dependences of the observed magnetic form factors at $h01$, $1k1$, $h03$, and $10l$ reflections 
estimated by using above $\phi$, $w_{yz}$ and $\mu$ values, are shown by closed circles, small open circles, 
closed triangles and large open circles, respectively. The $Q$-dependence of $f_{obs}(\bm{Q})$ 
at $1k1$ almost correspond with that of $10l$ reflections.  The decreases with $Q$ of $f_{obs}(\bm{Q})$ at 
$h01$ and $h03$ reflections are more gradual than those at $1k1$ and $10l$ reflections.  The lines which show the 
magnetic form factors calculated for above reciprocal lattice points, almost reproduce the $f_{obs}(\bm{Q})$.  
From these analyses, we know that the magnetic form factor has in-plane anisotropy.  
\par
The early neutron diffraction study on SrFe$_2$As$_2$ has claimed that magnetic form factor is approximately 
isotropic,\cite{ratcliff} inconsistent with our result.  In their data, the magnetic form factor at 501 reflection 
is zero.  We speculate that clear magnetic Bragg peak of the $501$ reflection can not be detected in their experimental accuracy 
because the structure factor of the $501$ reflection is much smaller than the intensities at the reciprocal lattice points with 
larger $l$ values.  If the $501$ reflection can be detected, their magnetic form factor must have similar in-plane 
anisotropy to our result, because the decrease with $Q$ of the magnetic form factor at $h01$ with $h$=1 and 3 is also slower 
than that at the other reciprocal lattice points, for example, $10l$ in their data.  
\par
The orbital ordering model in which the electrons in half-filled 3$d_{yz}$ orbital mainly gives the magnetic 
moment and the electrons in fully filled 3$d_{zx}$ orbital is non-magnetic, is suggested.\cite{lee}  
In such orbital ordering state, the antiferromagnetic magnetic interaction along $a$ axis is 
much larger than the interaction along $b$ axis, reproducing the stripe type magnetic structure and the large 
anisotropy of the spin wave dispersion.  
Schematic spin density expected from obtained magnetic form factor is shown in Fig. 4(d).  
The spin density is qualitatively consistent with the orbital ordering model although it is moderate relative to the above model.  
Our result can lead to the understanding the in-plane anisotropy of the magnetic behaviors.  
\par
In summary, the magnetic structure factor and the magnetic form factor which are determined by the neutron diffraction measurement 
on a detwinned single crystal of BaFe$_2$As$_2$ can be explained by considering that about half of the magnetic moment are contributed from the electrons in 3$d_{yz}$ orbital.  Such anisotropic spin density in $a$-$b$ plane is qualitatively consistent with the in-plane 
anisotropy of the magnetic behaviors observed in the antiferromagnetic phase of the parent compounds of the iron based superconductors.  
\par
This work was supported by a Grant-in-Aid for Specially Promoted Research 17001001 from 
the Ministry of Education, Culture, Sports, Science and Technology, Japan, and JST TRIP.  
This work was supported by a Grant-in-Aid for Scientific Research B (No. 24340090) from the Japan Society for the Promotion of Science.

\end{document}